\documentclass{jfm}

\usepackage{graphicx}
\usepackage{amsmath}

\providecommand{\backsection}[2][]{\par\smallskip\noindent\textbf{#1.}\enspace #2\par}

\newcommand{\cmin}{c_{\min}}
\newcommand{\wc}{\omega_c}
\newcommand{\kc}{k_c}
\newcommand{\Rmean}{\langle R\rangle}
\newcommand{\pe}{\hat{p}_e}
\newcommand{\pp}{p_e}              % real-space profile (no hat); \pe is its Fourier amplitude
\newcommand{\Pmag}{p}             % source amplitude (net force); line source: \pp = \Pmag\,\delta
\newcommand{\pfield}{\mathcal{P}}  % full applied pressure field P(x,t) = \pp(x+Vt)cos(Omega t)
\newcommand{\Vt}{\tilde{V}}        % V / cmin
\newcommand{\Ot}{\tilde{\Omega}}   % Omega / omega_c
\newcommand{\kt}{\tilde{k}}        % k / kc
\newcommand{\Rt}{\langle\tilde{R}\rangle}  % dimensionless mean resistance = <R>/(p^2/gamma), see (3.5)

\newif\ifdraftstamp \draftstampfalse
\newcommand{\tw}{\tilde\omega}
\newcommand{\ku}{\kt_{\mathrm{u}}}
\newcommand{\kd}{\kt_{\mathrm{d}}}    % racine double a une transition
\newcommand{\Od}{\Ot_{\mathrm{d}}}    % frequence de la transition
\newcommand{\draftversion}{v111}
\newcount\cmphh \newcount\cmpmm \newcount\cmptmp
\newcommand{\localtime}{\cmphh=\time \divide\cmphh by 60
  \cmptmp=\cmphh \multiply\cmptmp by 60 \cmpmm=\time \advance\cmpmm by -\cmptmp
  \ifnum\cmphh<10 0\fi\number\cmphh:\ifnum\cmpmm<10 0\fi\number\cmpmm}
\makeatletter
\ifdraftstamp
\ifnum\pdfshellescape>0
  \immediate\write18{TZ='Europe/Paris' date
    "+\@percentchar H:\@percentchar M (Paris)" > \jobname-tstamp.tex 2>/dev/null}
  \newcommand{\compiletime}{\input{\jobname-tstamp.tex}\unskip}
\else
  \newcommand{\compiletime}{\localtime\ (machine local)}
\fi
\fi
\makeatother

\title[Pulsation-induced wave resistance]
{Capillary--gravity wave resistance of a translating, pulsating source}

\author{Haim Diamant\aff{1}\corresp{\email{hdiamant@tau.ac.il}}
  \and Elie Raphael\aff{2}\corresp{\email{elie.raphael@espci.psl.eu}}}
\affiliation{\aff{1}School of Chemistry and Center for Physics and Chemistry of Living Systems, Tel Aviv University, 69978 Tel Aviv, Israel
  \aff{2}UMR CNRS Gulliver 7083, ESPCI Paris, PSL Research University, 75005 Paris, France}

\begin{document}

\maketitle

% --- ligne d'horodatage (PROVISOIRE, pilotee par \ifdraftstamp) ---
\ifdraftstamp
  \begin{center}\footnotesize\itshape
  Draft \draftversion\ --- compiled \today\ at \compiletime
  \end{center}
\fi

%------------------------------------------------------------------------------
\begin{abstract}
A disturbance in steady rectilinear translation along a capillary--gravity
interface radiates waves only above a minimum speed $\cmin$, where it
experiences a wave resistance; below this threshold it generates no waves, and the
response is merely a localised deformation of the
interface around the source, free of wave resistance. Modulating the source amplitude at a frequency $\Omega$ shifts
the forcing to the sideband pair $kV\pm\Omega$, where $k$ is the wavenumber and $V$ is the translation speed, opening resonance conditions
absent from the steady problem; radiation then sets in below the steady threshold $\cmin$. Here we
derive the mean wave resistance $\Rmean$ of a line source in closed
form, and show it to be strictly positive at every nonzero speed $V$: the drag-free
below-threshold regime disappears for any $\Omega\neq0$. Crossing the radiation
transitions---where the group velocity matches the source speed---produces integrable
resonant peaks in $\Rmean$, whose exponents we derive, including at the cusp where two
transition branches merge. Below threshold, where a steadily moving source feels no
drag, $\Rmean$ rises from zero as $\Omega^{5}$ (at fixed $V$). Throughout, surface tension is essential to $\Rmean$:
because it controls the short, dominant capillary-scale waves, discarding it misrepresents
$\Rmean$ over almost the entire speed--frequency plane.
\end{abstract}

% Mots-cles (liste officielle JFM).

%==============================================================================
\section{Introduction}
%==============================================================================
A disturbance moving along a fluid interface can radiate waves and, in doing so,
experience a resistive force, the \emph{wave resistance}.
This concept was developed by \citet{Havelock1918} for gravity waves on deep
water, in the wake of the classical studies of ship waves by
\citet{Kelvin1887} and the towing experiments of \citet{Froude1874}.
This Rapid article concerns the one-dimensional geometry of a line disturbance on a capillary--gravity interface
of an inviscid fluid.
For a source translating steadily at constant speed $V$ in this geometry, the
wave resistance has been calculated explicitly by \citet{Raphael1996} (hereafter RdG):
\begin{equation}
  R=
  \begin{cases}
    \dfrac{\Pmag^{2}}{\gamma}\left[1-(\cmin/V)^{4}\right]^{-1/2}, & V>\cmin,\\[1.5ex]
    0, & 0\leq V<\cmin,
  \end{cases}
  \label{eq:RdG}
\end{equation}
where $\Pmag$ is the magnitude of the localised pressure source and
$\cmin=(4g\gamma/\rho)^{1/4}$ is the minimum phase speed, set by the gravity $g$,
surface tension $\gamma$ and density $\rho$ of the liquid. Below $\cmin$ the
source radiates no waves, and the response is a deformation of the
interface localised around the source.

Many interfacial sources of practical interest are, however, not steady~\citep{Dode2022,Arutkin2026}. A
floating body that bobs while advancing, or a vertically vibrating probe scanning the
surface, imposes a pressure field whose \emph{amplitude} is modulated at some
frequency $\Omega$.
A source that translates uniformly while pulsating is a classical
problem of dispersive wave radiation~\citep{Lighthill1967}; for pure gravity waves a
resonance occurs at $\Omega V/g=\tfrac14$ \citep{Kaplan1957,DaganMiloh1982}, and in the gravity-free
limit its capillary counterpart is a critical resonance at
$\Omega\gamma/(\rho V^3)=4/27$ (\S\,\ref{sec:conclusion}). When both act together, by contrast,
the phase speed has a minimum, $\cmin$, which sets
the below-threshold regime of~\eqref{eq:RdG}, where $R=0$ for a source translating steadily. The
radiated wave field of the corresponding pulsating problem was obtained by \citet{Pramanik1980}: the radiating
wavenumbers are the roots of two cubics, transitions occur where the group velocity matches
the source speed, and the critical curves so defined partition the speed--frequency quadrant
into regions of different wave content, surface tension adding two capillary waves upstream. That study did not address the wave resistance, which is our subject here.
Unsteady motion is known to lift the below-threshold suppression: an accelerated
source---one that changes speed or direction---radiates at $V<\cmin$
\citep{Chepelianskii2008,Closa2010}, as does a laterally oscillating one.
\citet{RoccuzzoHerault2026} recently modelled the wake of a slaloming swimmer as a
superposition of translating pulsating sources, and found a wave-drag peak
below $\cmin$. Here, by contrast, the trajectory is the uniform rectilinear translation of
\citet{Raphael1996}, with no acceleration of any kind, and the spectral content is supplied
by amplitude modulation alone.

In this Rapid article we compute the wave resistance of this line source exactly. An amplitude
modulation $\cos(\Omega t)$ of a source translating at speed $V$ displaces the forcing to the two
sidebands $kV\pm\Omega$, opening resonance conditions with no counterpart in
the steady problem \citep{Pramanik1980}. We obtain $\Rmean$ in closed form, by contour
integration, as a finite sum over the radiating modes, and show it to be strictly positive at
\emph{every} speed $V>0$ as soon as $\Omega\neq0$ --- including the whole
below-threshold range where the steady source feels no wave resistance. Across the radiation
transitions $\Rmean$ develops integrable resonant singularities whose exponents we derive,
including at the cusp of one transition curve, where two saddle-node branches merge. The steady RdG result is
recovered continuously as $\Omega\to0$. Surface tension, finally, proves essential to
$\Rmean$ itself and not only to the existence of $\cmin$.

%==============================================================================
\section{Formulation and stationary response}\label{sec:formulation}
%==============================================================================
We consider a vertical pressure source $\pfield(x,t)$ translating at constant speed $V$ along the
interface of a deep, inviscid, irrotational fluid of density $\rho$, surface
tension $\gamma$ and gravity $g$.
The unperturbed interface lies on the $z=0$ plane, and the source moves along the $x$-axis, its centre at $x=-Vt$ (with $V\ge 0$). Its
amplitude is modulated harmonically at frequency $\Omega$,
\begin{equation}
  \pfield(x,t)=\pp(x+Vt)\,\cos(\Omega t).
  \label{eq:source}
\end{equation}
The source is invariant along the transverse $y$-direction, so the response is two-dimensional, with
surface elevation $\zeta(x,t)$. Expanding the $x$-dependence in Fourier modes $\zeta(x,t)=\int_{-\infty}^{\infty}\frac{\mathrm{d}k}{2\pi}\,\mathrm{e}^{\mathrm{i}kx}\,\hat\zeta(k,t)$,
the wavenumber $k$ ranges over all real values; a free wave of wavenumber $k$
oscillates at the frequency $\omega(k)=\sqrt{g|k|+(\gamma/\rho)|k|^3}$ (which depends only on $|k|$)~\citep{Lighthill1978}.
Since $\zeta$ is real, $\hat\zeta(-k,t)=\hat\zeta^*(k,t)$, and it suffices to take
$k>0$, for which $\omega(k)=\sqrt{gk(1+k^2/\kc^2)}$ is the
capillary--gravity dispersion relation, where $\kc=\sqrt{\rho g/\gamma}$. The phase speed $c(k)\equiv\omega(k)/k$ is minimal at the capillary wavenumber $\kc$,
where it equals $\cmin$, and $\wc=\cmin\kc=\sqrt{2g\kc}$; the group velocity is $c_g(k)\equiv\mathrm{d}\omega/\mathrm{d}k$. We
non-dimensionalise wavenumbers by $\kc$, speeds by $\cmin$ and frequencies by
$\wc$, writing $\kt=k/\kc$, $\Vt=V/\cmin$ and $\Ot=\Omega/\wc$; the
dispersion relation then reads $\omega(\kt)/\wc=\sqrt{(\kt+\kt^3)/2}$,
and the phase speed $c/\cmin=\sqrt{(1+\kt^2)/(2\kt)}$ is minimal at
$\kt=1$. The forcing of each surface mode is set by the spatial Fourier transform of
the pressure distribution $\pp$, $\pe(k)=\int_{-\infty}^{\infty}\pp(x)\,\mathrm{e}^{-\mathrm{i}kx}\,\mathrm{d}x$. For definiteness we take this distribution concentrated along a line, $\pp(x)=\Pmag\,\delta(x)$. Its Fourier transform is then constant, $\pe(k)=\Pmag$.

Linearising the equation of motion and the boundary condition at the free
surface, and eliminating the velocity potential \citep{Lighthill1978}, $\zeta(x,t)$ obeys, in Fourier space, the
forced-oscillator equation
\begin{equation}
  \partial_{tt}\hat\zeta(k,t)+\omega^2(k)\,\hat\zeta(k,t)
  =-\frac{k\,\Pmag}{2\rho}\bigl(\mathrm{e}^{\mathrm{i}\omega_{f1}t}
  +\mathrm{e}^{\mathrm{i}\omega_{f2}t}\bigr),
  \label{eq:oscillator}
\end{equation}
where the cosine modulation has split the forcing into two sidebands of the
Doppler frequency $kV$,
\begin{equation}
  \omega_{f1}=kV+\Omega,\qquad \omega_{f2}=kV-\Omega.
  \label{eq:forcing}
\end{equation}
The factor $k$ on the right-hand side of \eqref{eq:oscillator} reflects the non-local,
deep-fluid surface dynamics. Note that for a general pressure profile $\pp(x)$ of
finite extent, the constant amplitude $\Pmag$ on the right-hand side of
\eqref{eq:oscillator} is replaced by the Fourier transform $\pe(k)$.

The long-time solution of~\eqref{eq:oscillator} is, for $k>0$,
\begin{equation}
  \hat\zeta_c(k,t)=\frac{k\,\Pmag}{2\rho}
  \left[\frac{\mathrm{e}^{\mathrm{i}\omega_{f1}t}}{D_1-2\mathrm{i}\omega_{f1}\varepsilon}
  +\frac{\mathrm{e}^{\mathrm{i}\omega_{f2}t}}{D_2-2\mathrm{i}\omega_{f2}\varepsilon}\right],
  \qquad D_j=\omega_{fj}^2-\omega^2 ,
  \label{eq:stationary}
\end{equation}
where the Sommerfeld radiation condition is enforced by adding an infinitesimal
friction, $2\varepsilon\,\partial_t\hat\zeta$, to the left-hand side
of~\eqref{eq:oscillator}~\citep{Lighthill1978}.
Each radiating mode ($k>0$) appears as $\mathrm{e}^{\mathrm{i}(kx+\omega_{fj}t)}$,
a wave whose crests travel at the phase velocity $-\omega_{fj}/k$. As the source
advances in the $-x$ direction ($x=-Vt$), the sign of $\omega_f$ sets which way
these crests travel: $\omega_f>0$ moves them in the same direction as the source
($-x$; \emph{prograde}), $\omega_f<0$ in the opposite direction ($+x$;
\emph{retrograde}).

\begin{figure}
  \centering
  \includegraphics[width=\textwidth]{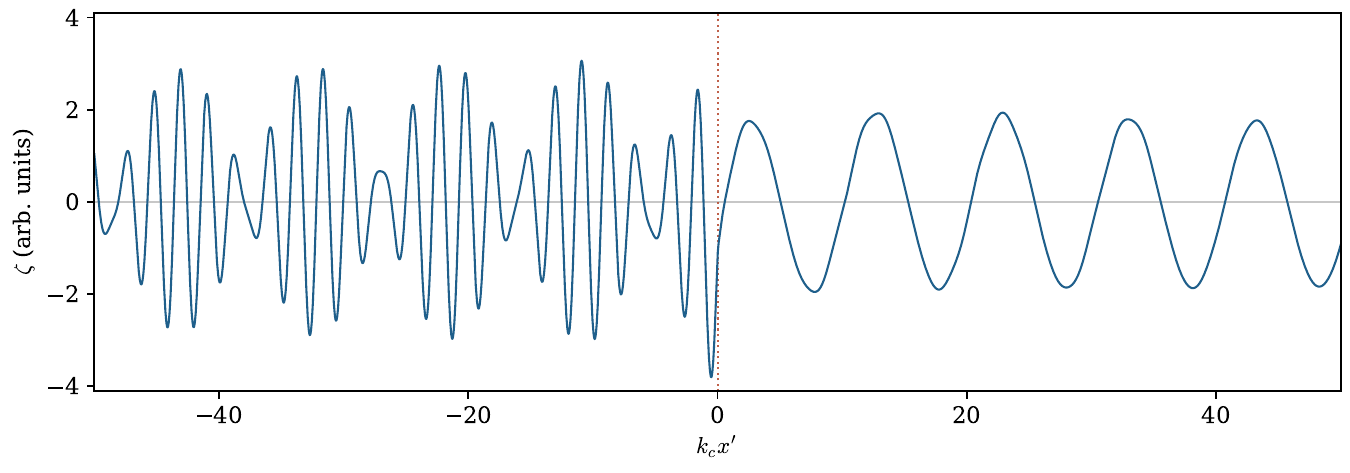}
  \caption{Interface profile $\zeta(x')$ in the source frame ($x'=x+Vt$) at $t/T=0$,
  $T=2\pi/\Omega$, from~\eqref{eq:stationary}, for $\Vt=1.3$, $\Ot=0.15$: two short capillary
  ripples ($\kt\approx2.7$ and $3.3$) on one side; a longer wave ($\kt\approx0.62$) with a
  slowly varying long-wave mode ($\kt\approx0.03$) on the other. The source is located at $x'=0$ (dotted line).}
  \label{fig:profile}
\end{figure}

The radiating modes are the real positive roots of $D_j$; in dimensionless form
$D_j\propto-P_j(\kt)$, introducing the two cubic polynomials that organise
the whole analysis,
\begin{align}
  P_1(\kt)&=\kt^3-2\Vt^2\kt^2+(1-4\Vt\Ot)\,\kt-2\Ot^2,\label{eq:P1}\\
  P_2(\kt)&=\kt^3-2\Vt^2\kt^2+(1+4\Vt\Ot)\,\kt-2\Ot^2.\label{eq:P2}
\end{align}
Three of their roots recur below: $\ku$ and $\kt_\ast$, the smallest positive root of
$P_1$ and the retrograde root of $P_2$ (both long-wavelength modes, $\sim\Ot^2$), and $\kt_{\rm d}$,
a double root, one at each radiation transition.
The damping term in~(\ref{eq:stationary}) is $-2\mathrm{i}\omega_{fj}\varepsilon$,
so the radiation condition enters with the sign of $\omega_{fj}$: negative for
the upper sideband $\omega_{f1}=kV+\Omega$, and reversed for the lower one
$\omega_{f2}=kV-\Omega$ once $\Omega>kV$.
The profile of the fluid interface as a function of the velocity and the
modulation frequency was first studied by \citet{Pramanik1980}; a representative
profile is shown in figure~\ref{fig:profile}.
How many of these roots are real and positive---and hence which waves
radiate---depends on $(\Vt,\Ot)$; the resulting partition of the plane into
four radiation regimes is the phase diagram of figure~\ref{fig:phase}
(\S\,\ref{sec:phasediagram}).

%==============================================================================
\section{Mean wave resistance}\label{sec:resistance}
%==============================================================================
The wave resistance is the streamwise force exerted by the fluid on the source~\citep{Havelock1918},
reckoned positive when it opposes the motion,
\begin{equation}
  R(t)=-\int_{-\infty}^{+\infty}\pfield(x,t)\,\partial_x\zeta(x,t)\,\mathrm{d}x ,
  \label{eq:Rdef}
\end{equation}
where $\pfield(x,t)$ is given by \eqref{eq:source}.
Using Parseval's theorem, we substitute \eqref{eq:stationary} and average over
a modulation period; only the $\cos^2\Omega t$ contributions survive, giving
\begin{equation}
  \Rmean=\int_0^\infty\frac{k^2\Pmag^2}{\rho}
  \left[\frac{\omega_{f1}\varepsilon}{D_1^2+4\omega_{f1}^2\varepsilon^2}
  +\frac{\omega_{f2}\varepsilon}{D_2^2+4\omega_{f2}^2\varepsilon^2}\right]\frac{\mathrm{d}k}{2\pi}.
  \label{eq:Rintegral}
\end{equation}
In the radiation limit $\varepsilon\to0^+$ each Lorentzian collapses onto a
signed delta, $\omega_f\varepsilon/(D^2+4\omega_f^2\varepsilon^2)\to
(\pi/2)\,\mathrm{sgn}(\omega_f)\,\delta(D)$, and the integral reduces to a sum of
residues over the roots of $P_1$ and $P_2$. In terms of the dimensionless
wavenumber $\kt=k/\kc$, this gives
\begin{equation}
  \Rt\equiv\frac{\Rmean}{\Pmag^2/\gamma}
  =\frac{1}{4}\bigl(\tilde S_1+\tilde S_2^{>}-\tilde S_2^{<}\bigr),
  \label{eq:Rformula}
\end{equation}
where
\begin{equation}
  \tilde S_1=\!\!\sum_{\substack{P_1(\kt_n)=0\\ \kt_n>0}}\,
  \frac{\kt_n^2}{|P_1'(\kt_n)|},
  \qquad
  \tilde S_2^{\gtrless}=\!\!\sum_{\substack{P_2(\kt_n)=0\\ \kt_n\Vt\gtrless\Ot}}\,
  \frac{\kt_n^2}{|P_2'(\kt_n)|}.
  \label{eq:S}
\end{equation}
The three sums in~\eqref{eq:S} are non-negative; they differ only by the sign with
which $\mathrm{sgn}(\omega_{fj})$ makes them enter~\eqref{eq:Rformula}. The upper sideband
has $\omega_{f1}=kV+\Omega>0$ at every $k>0$, so $\tilde S_1$ always enters with a plus;
of the lower-sideband waves, the prograde ones ($\omega_{f2}>0$) enter with a plus and the
retrograde ones ($\omega_{f2}<0$) with a minus. Writing $\tw(\kt)=\omega/\wc$, one has
$P_j=2[\tw^2-(\kt\Vt\pm\Ot)^2]$, so the positive roots of $P_1$ solve
$\tw-\kt\Vt=\Ot$, and those of $P_2$ solve $\tw-\kt\Vt=-\Ot$ (prograde) or
$\tw+\kt\Vt=\Ot$ (retrograde). The last equation has exactly one root $\kt_\ast$,
its left-hand side rising monotonically from $0$ to $\infty$: there is always exactly
one retrograde wave.

The three sums combine in~\eqref{eq:Rformula} with $\tilde S_2^{<}$ subtracted, so
$\Rmean$ is not manifestly positive. It is nevertheless strictly positive for every $\Vt>0$
and $\Omega\neq0$. To prove this, let $\ku$ be the smallest positive root of $P_1$, which exists since
$P_1(0)=-2\Ot^2<0$. Because $\tw+\kt\Vt>\tw-\kt\Vt$ for $\kt>0$, the retrograde root $\kt_\ast$ lies
below $\ku$; and the identity $\tw\,c_g/\cmin=(1+3\kt^2)/4$ then gives
$\ku^2/P_1'(\ku)>\kt_\ast^2/P_2'(\kt_\ast)$.\footnote{This inequality follows from the derivatives: with
$\tw_{\mathrm{u}}=\ku\Vt+\Ot$ and $\tw_\ast=\Ot-\kt_\ast\Vt$ the dispersion at the
two roots, both positive, $P_1'(\ku)=1+3\ku^2-4\Vt\tw_{\mathrm{u}}$ and
$P_2'(\kt_\ast)=1+3\kt_\ast^2+4\Vt\tw_\ast$. The second is positive term by term; the first
is non-negative, $\ku$ being the first root of $P_1$, and vanishes only at a radiation
transition, where $\Rmean$ diverges. Except at a transition, both denominators are
positive, and the inequality is equivalent to $\ku^2P_2'(\kt_\ast)-\kt_\ast^2P_1'(\ku)>0$. In
this difference the $3\ku^2\kt_\ast^2$ terms cancel, leaving
$\ku^2-\kt_\ast^2+4\Vt(\ku^2\tw_\ast+\kt_\ast^2\tw_{\mathrm{u}})$, positive since
$\kt_\ast<\ku$.} The left-hand side is one term of $\tilde S_1$ and the
right-hand side is all of $\tilde S_2^{<}$, so $\tilde S_1>\tilde S_2^{<}$: the single
upper-sideband wave $\ku$ already outweighs the retrograde one $\kt_\ast$, which therefore cannot
reverse the balance. We conclude that the mean wave resistance $\Rmean$ is always positive, and
vanishes only on the steady
below-threshold axis $\Omega=0$, $\Vt<1$, where nothing is radiated, and on the axis
$\Vt=0$, where a source that does not advance radiates symmetrically. Note that the sign
of the instantaneous wave resistance
$R(t)=\Rmean(1+\cos2\Omega t)-C\sin2\Omega t$ (with $C$ the reactive, non-radiating
amplitude) varies in time, dipping to $\Rmean-\sqrt{\Rmean^2+C^2}<0$ within each period.

%==============================================================================
\section{Below-threshold radiation and the phase diagram}\label{sec:phasediagram}
%==============================================================================
The central consequence of \eqref{eq:Rformula} is immediate. For any
$\Omega\neq0$ the small root $\kt_\ast$ of $P_2$ exists at \emph{any} speed
$V>0$: it represents a genuinely propagating mode, radiating power to infinity.
It is retrograde ($\omega_{f2}<0$) and thus enters the residue
sum~\eqref{eq:Rformula} through the subtracted term $\tilde S_2^{<}$. The upper
sideband is never silent either: $P_1(0)=-2\Ot^{2}<0$ guarantees a positive root $\ku$
of $P_1$ for every $\Omega\neq0$, and by the inequality of~\S\,\ref{sec:resistance}
that single wave already outweighs the retrograde one $\kt_\ast$, so that $\Rmean>0$. The
steady threshold $V=\cmin$ therefore does not carry over to the modulated
problem --- as soon as the source oscillates, radiation occurs at all speeds,
and the below-threshold silent regime of the steady source disappears. Whirligig
beetles radiate capillary waves below $\cmin$ through their leg strokes
\citep{VoiseCasas2010}, an effect ascribed to the unsteadiness of the strokes
\citep{Closa2010}; their periodicity, however, is by itself enough.

\begin{figure}
  \centering
  \includegraphics[width=0.78\textwidth]{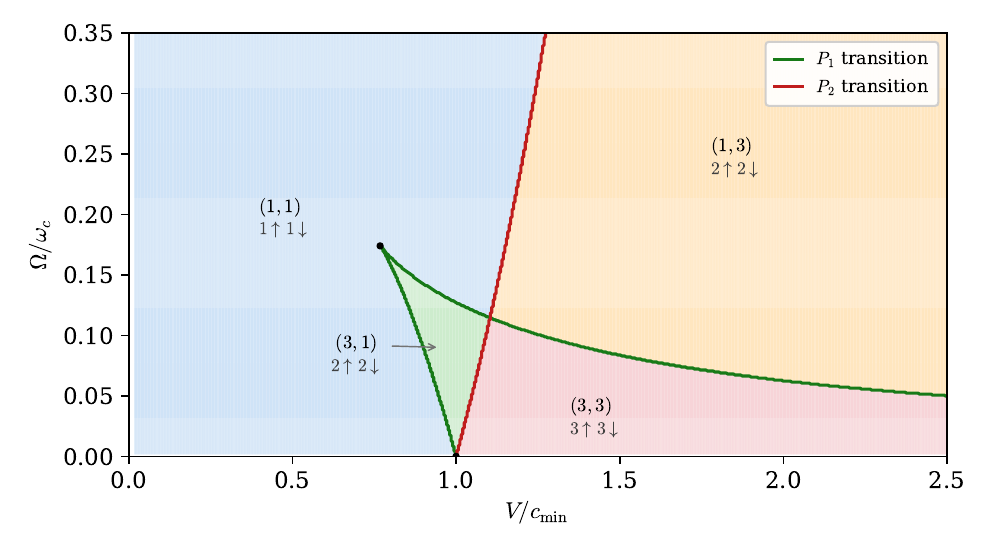}
  \caption{Phase diagram in the $(\Vt,\Ot)$ plane. Each region is labelled by
  the numbers $(n_1,n_2)$ of real positive roots of $(P_1,P_2)$ and, beneath, by
  the split of the radiating modes into upstream ($\uparrow$, energy ahead of
  the source) and downstream ($\downarrow$, behind); the two counts are equal in
  every region (the narrow $(3,1)$ region is labelled by a leader). The green
  and red curves are the radiation transitions of $P_1$ and $P_2$;
  both emanate from the bifurcation point $(\Vt,\Ot)=(1,0)$, and the cusp of the
  $P_1$ curve lies at $(\Vt,\Ot)\approx(0.768,0.174)$. The expression for
  $\Rmean$ in each region is given by \eqref{eq:Rformula}.}
  \label{fig:phase}
\end{figure}

The number of radiating modes changes across \emph{radiation transitions},
at which two roots of $P_j$ coalesce into a double root $\kd$. This is the condition
$c_g(\kd)=V$ on the group velocity (i.e. $c_g/\cmin=\Vt$), which traces
one curve for each polynomial in the
$(\Vt,\Ot)$ plane and partitions it into the four regions of the phase
diagram, figure~\ref{fig:phase}. The criterion, and the
partition it induces --- including the cusp below --- were obtained by \citet{Pramanik1980},
within the general framework of \citet{Lighthill1967}. Both transition curves emanate from the steady
bifurcation point $(\Vt,\Ot)=(1,0)$; the $P_1$ curve possesses a cusp at
$\kt_{\rm f}=\sqrt{(2\sqrt3-3)/3}\approx0.393$, with coordinates
$\Vt_{\rm f}=(3-\sqrt3)/[2(2\sqrt3-3)^{1/4}]\approx0.768$ and
$\Ot_{\rm f}=(2\sqrt3-3)^{1/4}(3-\sqrt3)/6\approx0.174$.

The same group-velocity criterion fixes \emph{where} each radiated wave is found
relative to the source. A mode sits upstream (ahead) when its energy---carried at
the group velocity---outruns the source, and downstream (behind) otherwise. A
prograde mode ($\omega_{fj}>0$) carries its energy forwards, so it sits upstream
when $c_g>V$, i.e.\ when $P_j'(\kt_n)>0$. A retrograde mode
($\omega_{f2}<0$) carries its energy backwards and is therefore always
downstream. This generalises the steady picture, in which
the short capillary wave rides ahead of the source while the long gravity wave
trails behind. Each sideband favours one side: the upper sideband ($\omega_{f1}$)
always supplies one net extra upstream mode --- at small $\Ot$, its long prograde wave
$\ku\sim\Ot^2$, with $c_g\gg V$ --- whereas the lower sideband ($\omega_{f2}$) supplies one net extra downstream mode, the retrograde long wave $\kt_\ast$
($\omega_{f2}<0$). The radiating modes therefore split into equal numbers
upstream and downstream in every region, as annotated in
figure~\ref{fig:phase}. This balance has no counterpart in the pure-gravity
problem, where one wave travels upstream and three downstream below the
resonance $\Omega V/g=\tfrac14$, only two downstream waves remaining above it
\citep{Kaplan1957,DaganMiloh1982}. Surface tension adds two capillary waves, both
upstream \citep{Pramanik1980}.

The transitions leave a clear signature in $\Rmean$. At a double root,
$P_j'(\kd)=0$ and the term $\kd^2/|P_j'(\kd)|$ diverges, so that
$\Rmean$ develops an integrable singularity.

\begin{figure}
  \centering
  \includegraphics[width=\textwidth]{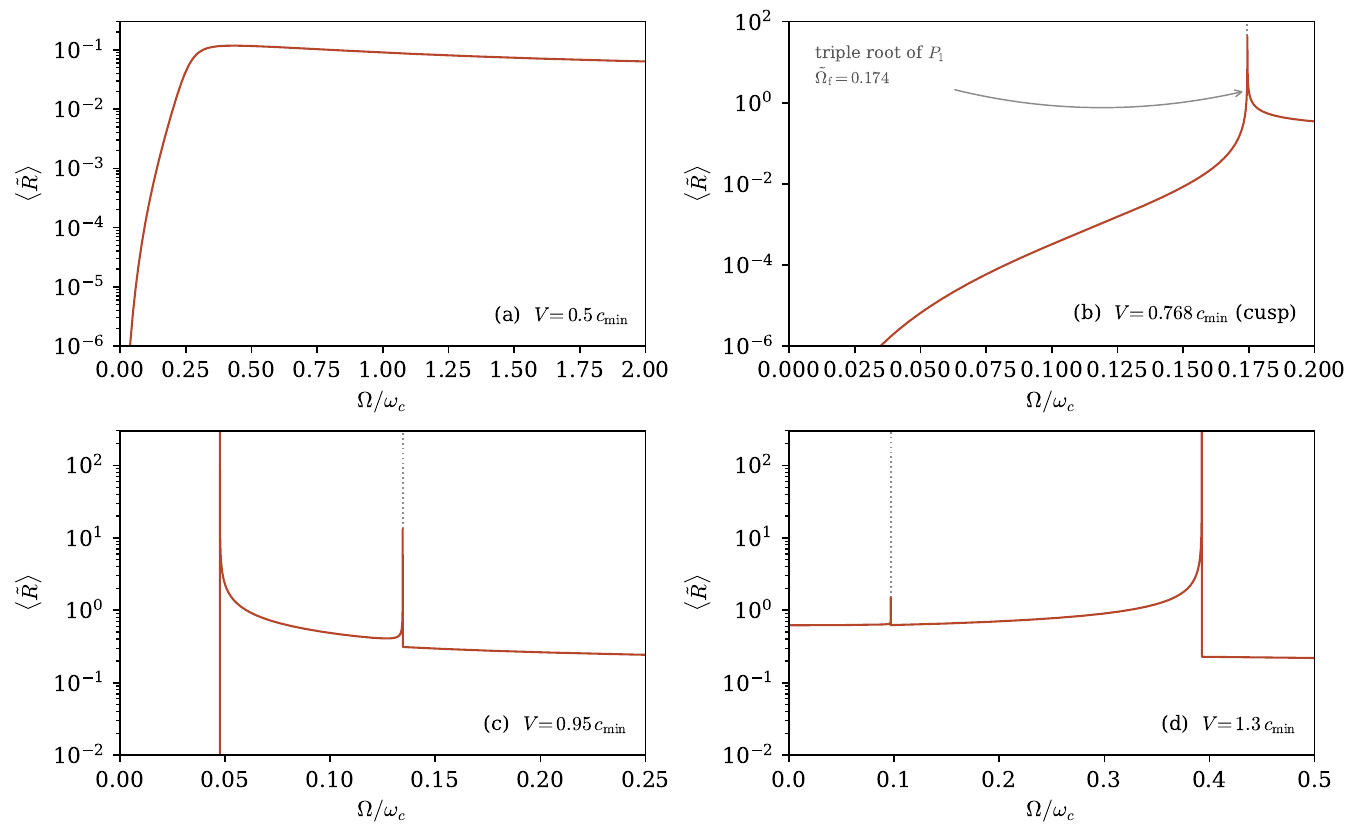}
  \caption{Dimensionless mean wave resistance $\Rt=\Rmean/(\Pmag^2/\gamma)$ (logarithmic
  scale) versus $\Ot=\Omega/\wc$ at four fixed speeds, sweeping $V$ through the $P_1$ cusp
  and the steady threshold. Throughout, $\Rt$ \emph{diverges integrably at the radiation
  transitions} (dotted asymptotes), where a radiating root of $P_1$ or $P_2$ becomes double,
  and stays finite elsewhere; it decays as $\Ot^{-1/3}$ at high frequency.
  (\textit{a}) At $V=0.5\,\cmin$, below the cusp speed, the vertical cut meets no $P_1$
  transition: $\Rt$ rises as $\Ot^{5}$ at small $\Ot$ and passes a broad maximum, without any
  resonance. (\textit{b}) At the cusp speed $V=0.768\,\cmin$ the two $P_1$
  saddle-node branches meet and the radiating root becomes triple at $\Ot_{\rm f}=0.174$,
  where $\Rt$ diverges on \emph{both} sides as $|\Ot-\Ot_{\rm f}|^{-2/3}$.
  (\textit{c}) At $V=0.95\,\cmin$, although $V<\cmin$, the source radiates at every
  $\Omega\neq0$ ($\Rt$ vanishes only at $\Omega=0$ and is too small to be visible before the
  first transition), crossing two $P_1$ transitions ($\Ot\approx0.048,\,0.134$).
  (\textit{d}) Above threshold, $V=1.3\,\cmin$, the resistance starts from its finite steady
  value, crosses a low-amplitude $P_1$ resonance ($\Ot\approx0.097$) and a high-amplitude
  $P_2$ resonance ($\Ot\approx0.393$) --- both integrable divergences, of very different
  prefactor --- beyond which a radiating mode disappears and the resistance drops.}
  \label{fig:Rpanels}
\end{figure}

Close to a transition, reached at $\Ot=\Od$ for a given $\Vt$, the two coalescing
roots form a saddle-node, $\kt_\pm=\kd\pm\sqrt{\alpha\,(\Ot-\Od)}$, with the saddle-node coefficient $\alpha=-2\,\partial_{\Ot}P_j/P_j''$ evaluated at $(\kd,\Od)$, so that $|P_j'(\kt_\pm)|\propto
|\Ot-\Od|^{1/2}$ and the resistance diverges as an integrable, one-sided
square root, $\Rt\simeq A(\Vt)\,|\Ot-\Od|^{-1/2}$ on the side where the
pair is real (for instance $A\approx0.11$ at $\Vt=0.95$). This square-root law
degenerates at the bifurcation point $(\Vt,\Ot)=(1,0)$, where the coalescence is
pinned at the dispersion minimum $\kt=1$ and the two transition curves
originate. There the steady limit of $\Rt$ already diverges as
$\Rt\simeq\tfrac14(\Vt-1)^{-1/2}$ along $\Ot=0$, while along the threshold
$\Vt=1$ the modulated wave resistance diverges as $\Rt\simeq\tfrac18\,\Ot^{-1/2}$ as
$\Ot\to0$; the bifurcation point is thus the common origin of the steady and
modulated square-root singularities. At the cusp of the $P_1$ curve the two
saddle-node branches meet, the radiating root becoming triple, and the one-sided square root is replaced
by a stronger, two-sided divergence $\Rt\propto|\Ot-\Ot_{\rm f}|^{-2/3}$
(figure~\ref{fig:Rpanels}\textit{b}). Conversely, for $\Vt$ below the cusp speed a vertical cut
meets no $P_1$ transition and $\Rmean$ rises smoothly as $\Ot^{5}$ at small $\Ot$ to a broad maximum without
resonance, after which it decays as $\Ot^{-1/3}$ (figure~\ref{fig:Rpanels}\textit{a}).

For a speed just below the steady
threshold, a vertical cut at fixed $V$ crosses these transition curves and
$\Rmean$ exhibits sharp resonant peaks (figure~\ref{fig:Rpanels}). At the
below-threshold speed $V=0.95\,\cmin$ (figure~\ref{fig:Rpanels}\textit{c}) two such
peaks appear within $\Ot\lesssim0.2$: a first, high-amplitude peak near
$\Ot\approx0.048$ where two complex-conjugate roots of $P_1$ coalesce on the
real axis, and a second near $\Ot\approx0.134$ where the reverse
transition takes place. A slow temporal modulation of a
marginally below-threshold source is thus sufficient to trigger a pronounced
radiative response. Above threshold ($V=1.3\,\cmin$,
figure~\ref{fig:Rpanels}\textit{d}) the wave resistance instead starts from its finite
steady value, passes a low-amplitude $P_1$ resonance and a high-amplitude $P_2$
resonance, and drops as the latter transition removes a radiating mode.

%==============================================================================
\section{Limiting and asymptotic behaviour}\label{sec:limits}
%==============================================================================
\emph{Recovery of the steady limit.} As $\Omega\to0^+$ the two long sideband roots
$\ku$ and $\kt_\ast$ tend to zero and their contributions vanish, while the remaining roots of
$P_1$ and $P_2$ merge onto those of $P_0(\kt)=\kt^3-2\Vt^2\kt^2+\kt$.
The formula \eqref{eq:Rformula} then reduces to
$\Rt=\tfrac12\!\sum_{P_0(\kt_n)=0,\,\kt_n>0}\,
\kt_n^2/|P_0'(\kt_n)|$, which is zero for $\Vt<1$ (no real positive root
of $P_0$) and positive for $\Vt>1$. This is exactly one half of the steady
RdG resistance~\eqref{eq:RdG}: the factor $\tfrac12$ is the time-average
$\langle\cos^2\Omega t\rangle$ of the modulation, the source amplitude
$\Pmag\cos\Omega t$ entering the quadratic resistance through its mean square.
Up to this numerical factor, the modulated
result therefore connects continuously to the RdG resistance above the
threshold and to the wave-free, resistance-free state below it.

\emph{Below-threshold asymptotics.} In the wave-resistance-free corner of the steady problem
($V<\cmin$, small $\Omega$), the two long roots $\ku$ and $\kt_\ast$ are nearly
degenerate and their residues almost cancel; the leading non-zero contribution scales as
\begin{equation}
  \Rt\;\xrightarrow[\Ot\to0]{}\;24\,\Vt\,\Ot^{5}.
  \label{eq:smallOmega}
\end{equation}
At the opposite end the radiating wavenumbers grow without bound,
$\kt_n\sim\Ot^{2/3}$; at such large $\kt$ the capillary term $\kt^3$ dominates the
dispersion relation over the gravity term $\kt$, so the radiated waves are short,
pure capillary ripples. The order-$\Ot^0$ terms cancel, and
\begin{equation}
  \Rt\;\xrightarrow[\Ot\to\infty]{}\;
  \frac{2^{1/3}}{9}\,\Vt\,\Ot^{-1/3},
  \label{eq:largeOmega}
\end{equation}
with the universal coefficient $2^{1/3}/9\approx0.140$. Since $\cmin$ enters~(\ref{eq:largeOmega})
only through the scales, the high-frequency limit retains
no signature of the steady threshold: below- and above-threshold sources alike
approach the same capillary decay~(\ref{eq:largeOmega}),
$\Rmean\propto\Ot^{-1/3}$, differing only through the prefactor $\propto\Vt$. Below threshold this
raises $\Rmean$ from zero; above threshold it instead drives $\Rmean$ well below
the steady RdG value, so that rapid enough modulation \emph{suppresses}
the wave drag of an above-threshold source --- the exact opposite of the radiation it
\emph{induces} below threshold. Finally, for $V\to\infty$ at fixed
$\Omega$ the radiation is carried by short capillary ripples
($\kt_n\sim2\Vt^2$), and the resistance flattens onto a plateau set by surface
tension alone, $\Rt\to\tfrac12+\Ot/(4\Vt^{3})+3/(16\Vt^{4})$. The leading term
$\tfrac12$ --- half the steady RdG value $\Pmag^2/\gamma$ --- is once more the
$\langle\cos^2\Omega t\rangle$ factor, but that factor does not survive at higher
order: the $\Vt^{-4}$ coefficient is $3/16$, not the $1/4$ that factor would give. The expansion is therefore
non-analytic at $\Omega=0$, and the limits $V\to\infty$ and $\Omega\to0$ \emph{do not commute}.

\begin{figure}
  \centering
  \includegraphics[width=0.6\textwidth]{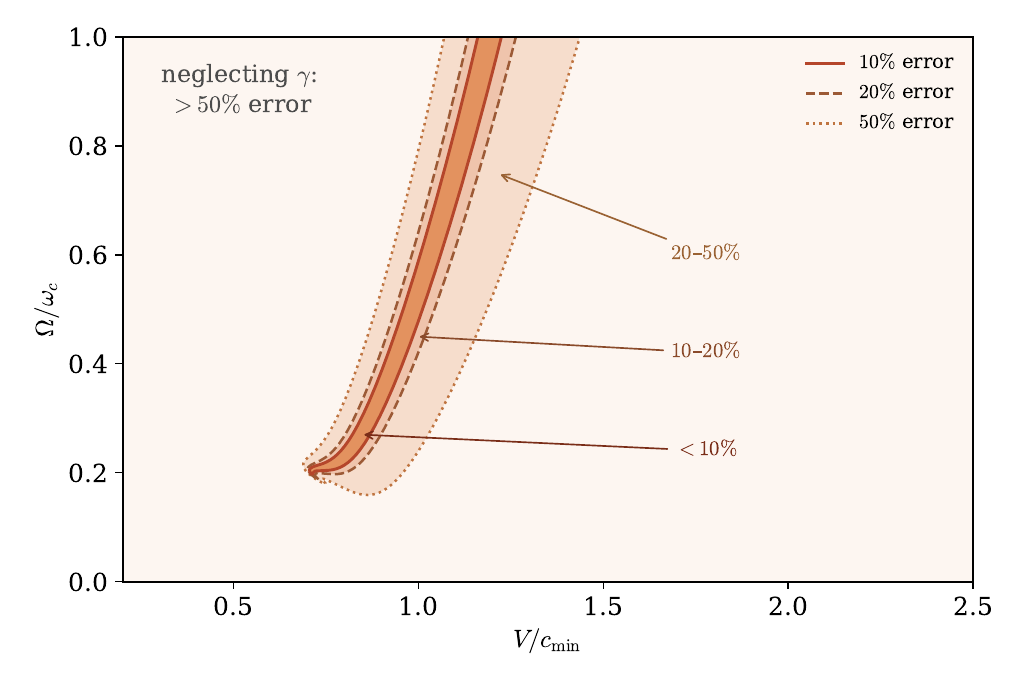}
  \caption{Failure of the pure-gravity approximation for the mean wave resistance
  $\Rmean$ of a line source, obtained by discarding surface tension (dispersion $\omega^2=gk$).
  Contours mark the relative error reaching $10\%$ (solid), $20\%$ (dashed) and
  $50\%$ (dotted): the approximation is accurate to within $10\%$ only along the
  thin band where its single gravity branch happens to mimic the exact response,
  while the error exceeds $50\%$ over most of the $(\Vt,\Ot)$ plane. Neglecting
  $\gamma$ thus misrepresents $\Rmean$ almost everywhere.}
  \label{fig:validity}
\end{figure}

Because each radiating mode contributes to $\Rmean$ in proportion to
$\kt_n^2$ (equation~\ref{eq:S}), the resistance is dominated by the short,
capillary-scale waves throughout most of the $(\Vt,\Ot)$ plane --- everywhere, that is,
except the below-threshold corner where the radiated waves are long. As a result the
two single-physics reductions --- approximating the dispersion relation by
capillarity alone or by gravity alone --- are far from equally accurate. To compare the two, we keep the full
physical system (gravity \emph{and} surface tension), so that
$\cmin$ and $\wc$ remain well defined and set the scales of
figure~\ref{fig:validity}, and merely simplify the dispersion relation used
to compute the radiated field. Retaining only the capillary
term ($\omega^2=\gamma k^3/\rho$) reproduces $\Rmean$ to within a few per cent wherever
the radiating waves are short, at high speed ($\Vt\gtrsim2$) or high frequency; but it has
no threshold, and below $\cmin$ at small $\Ot$ it misses the near-cancellation of residues,
so that by the analysis of \eqref{eq:smallOmega} it overestimates $\Rmean$ by orders of magnitude. Discarding surface
tension instead, that is using the pure-gravity dispersion $\omega^2=gk$,
misrepresents $\Rmean$ by more than $10\%$ over almost the entire $(\Vt,\Ot)$
plane (figure~\ref{fig:validity}): although $\gamma$ is dynamically negligible
for the long gravity waves, it controls the short, dominant capillary-scale
waves and therefore $\Rmean$ itself almost everywhere.

%==============================================================================
\section{Conclusion}\label{sec:conclusion}
%==============================================================================
We have shown that temporal modulation qualitatively changes the wave
resistance of a source moving along a capillary--gravity interface. The cosine
modulation shifts the forcing to the sideband pair $kV\pm\Omega$, and the resulting small
roots $\ku$ of $P_1$ and $\kt_\ast$ of $P_2$ guarantee propagating modes --- and, by the
inequality of \S\,\ref{sec:resistance}, a strictly positive mean wave resistance --- at
every speed $V>0$ once $\Omega\neq0$. The
steady radiation threshold $V=\cmin$ does not survive modulation: a slow
oscillation suffices to make a marginally below-threshold source radiate, with a
mean wave resistance that develops integrable resonant peaks where the
radiation transitions are crossed.

Physically, the below-threshold radiation takes the form of very long-wavelength
modes, $\ku\sim\kt_\ast\sim\Ot^2$ as $\Ot\to0$, superposed on the
localised deformation of the interface near the source. This tail leaves a striking
signature in the net displaced area. For a steady source the displaced volume is
finite, $\int_{-\infty}^{\infty}\zeta\,\mathrm{d}x=-\Pmag/(\rho g)$, a
long-wavelength hydrostatic balance \citep{Raphael1996, Lamb1932}. Under modulation
this static displacement gives way to a purely oscillatory one, whose displaced area
vanishes at every instant: $\int_{-\infty}^{\infty}\zeta\,\mathrm{d}x=0$
for every $\Omega\neq0$, a discontinuous change at $\Omega=0$. The resistance, by contrast, is recovered continuously as
$\Omega\to0$, reducing to the steady RdG result~\eqref{eq:RdG} up to the
time-average $\langle\cos^2\Omega t\rangle=\tfrac12$. In each of the two opposite single-scale limits
the response reduces to a single critical resonance: the surface-tension-free limit reproduces, within
linear theory, the critical resonance $\Omega V/g=\tfrac14$ of the classical
translating--pulsating gravity-wave source \citep{Kaplan1957,DaganMiloh1982}, while in the
gravity-free limit ($g\to0$) the present theory predicts its capillary
counterpart, a single critical condition $\Omega\gamma/(\rho V^3)=4/27$
(equivalently $\Omega=4\rho V^3/(27\gamma)$, with resonant wavenumber
$k_{\mathrm{d}}=4\rho V^2/(9\gamma)$), the capillary scaling $\Omega\propto V^3$
contrasting with the $\Omega\propto V^{-1}$ scaling of the gravity resonance.

The results are directly relevant to interfacial objects whose forcing, rather than
being steady, is periodically modulated as they translate, and they show how the wave
resistance is then to be computed. Two natural extensions suggest themselves. The first replaces the line source by a
source of finite width, which enters only through its Fourier transform $\pe(k)$
[cf.~\eqref{eq:oscillator}] and reshapes the radiated spectrum. The second carries
the time-dependent analysis over to a localised disturbance on the full
two-dimensional interface, generalising the steady result of
\citet[Eq.~(2.36)]{Raphael1996}. The present analysis is linearised throughout (small-amplitude waves):
while \citet{DaganMiloh1982} also addressed the nonlinear regime for pure
gravity waves, the nonlinear capillary--gravity~\citep{DiasKharif1999} problem remains open. Finally, the fluid has
been taken inviscid; viscosity is known to keep the steady wave resistance
bounded as $V\to\cmin$ \citep{RichardRaphael1999}, and assessing how it tempers
the resonant divergences of the modulated problem is a natural further direction.
Experimentally, such a forcing could be realised by a vertically oscillating
microprobe or AFM tip in tapping mode---whose lateral force directly measures
$\Rmean$---or, contactlessly, by an amplitude-modulated focused acoustic
radiation pressure.

%------------------------------------------------------------------------------
% Declarations exigees par JFM (a completer)
\backsection[Acknowledgements]{HD is grateful to the Gulliver laboratory, ESPCI Paris, for its hospitality.}
\backsection[Funding]{HD acknowledges support from the Israel Science Foundation (Grant No. 1611/24).}
\backsection[Declaration of interests]{The authors report no conflict of interest.}
\backsection[Data availability statement]{The scripts supporting this study are available from the corresponding authors on reasonable request.}

%==============================================================================
%  References (natbib / jfm) --- entrees renseignees et verifiees.
%  Style natbib (jfm). Completer avec un .bib + \bibliographystyle{jfm}.
%==============================================================================


\begin{thebibliography}{}

\bibitem[Arutkin et~al.(2026)Arutkin, Reuveni \& Raphael]{Arutkin2026}
  {\sc Arutkin, M., Reuveni, S. \& Raphael, E.} 2026
  Wave resistance for stochastic motion at interfaces.
  {\it Preprint}, arXiv:2606.08732.

\bibitem[Chepelianskii et~al.(2008)Chepelianskii, Chevy \& Raphael]{Chepelianskii2008}
  {\sc Chepelianskii, A. D., Chevy, F. \& Raphael, E.} 2008
  Capillary-gravity waves generated by a slow moving object.
  {\it Phys.\ Rev.\ Lett.} {\bf 100}, 074504.

\bibitem[Closa et~al.(2010)Closa, Chepelianskii \& Raphael]{Closa2010}
  {\sc Closa, F., Chepelianskii, A. D. \& Raphael, E.} 2010
  Capillary-gravity waves generated by a sudden object motion.
  {\it Phys.\ Fluids} {\bf 22}, 052107.

\bibitem[Dagan \& Miloh(1982)]{DaganMiloh1982}
  {\sc Dagan, G. \& Miloh, T.} 1982
  Free-surface flow past oscillating singularities at resonant frequency.
  {\it J.\ Fluid Mech.} {\bf 120}, 139--154.

\bibitem[Dias \& Kharif(1999)]{DiasKharif1999}
  {\sc Dias, F. \& Kharif, C.} 1999
  Nonlinear gravity and capillary-gravity waves.
  {\it Annu.\ Rev.\ Fluid Mech.} {\bf 31}, 301--346.

\bibitem[Dode et~al.(2022)Dode, Carmigniani, Cohen, Clanet \& Bocquet]{Dode2022}
  {\sc Dode, A., Carmigniani, R., Cohen, C., Clanet, C. \& Bocquet, L.} 2022
  Wave drag during an unsteady motion.
  {\it J.\ Fluid Mech.} {\bf 951}, A15.

\bibitem[Froude(1874)]{Froude1874}
  {\sc Froude, W.} 1874
  Experiments with H.M.S.\ Greyhound.
  {\it Trans.\ Inst.\ Nav.\ Archit.} {\bf 15}, 36--73.

\bibitem[Havelock(1918)]{Havelock1918}
  {\sc Havelock, T.~H.} 1918
  Wave resistance: some cases of three-dimensional fluid motion.
  {\it Proc.\ R.\ Soc.\ Lond.\ A} {\bf 95}, 354--365.

\bibitem[Kaplan(1957)]{Kaplan1957}
  {\sc Kaplan, P.} 1957
  The waves generated by the forward motion of oscillatory pressure distributions.
  In {\it Proc.\ 5th Midwestern Conf.\ Fluid Mech.}, pp.~316--329.

\bibitem[Lamb(1932)]{Lamb1932}
  {\sc Lamb, H.} 1932
  {\it Hydrodynamics}, 6th edn.
  Cambridge University Press. (Reprinted 1993.)

\bibitem[Lighthill(1967)]{Lighthill1967}
  {\sc Lighthill, M. J.} 1967
  On waves generated in dispersive systems by travelling forcing effects, with
  applications to the dynamics of rotating fluids.
  {\it J.\ Fluid Mech.} {\bf 27}, 725--752.

\bibitem[Lighthill(1978)]{Lighthill1978}
  {\sc Lighthill, M. J.} 1978
  {\it Waves in Fluids}.
  Cambridge University Press.

\bibitem[Pramanik(1980)]{Pramanik1980}
  {\sc Pramanik, A.~K.} 1980
  Capillary-gravity waves produced by a moving pressure distribution.
  {\it Z.\ Angew.\ Math.\ Phys.} {\bf 31}, 174.

\bibitem[Raphael \& de Gennes(1996)]{Raphael1996}
  {\sc Raphael, E. \& de Gennes, P.-G.} 1996
  Capillary gravity waves caused by a moving disturbance: wave resistance.
  {\it Phys.\ Rev.\ E} {\bf 53}, 3448--3455.

\bibitem[Richard \& Raphael(1999)]{RichardRaphael1999}
  {\sc Richard, D. \& Raphael, E.} 1999
  Capillary-gravity waves: the effect of viscosity on the wave resistance.
  {\it Europhys.\ Lett.} {\bf 49}, 53--58.

\bibitem[Roccuzzo \& Herault(2026)]{RoccuzzoHerault2026}
  {\sc Roccuzzo, M. \& Herault, J.} 2026
  Transitions in unsteady capillary-gravity wakes of surface swimmers.
  {\it Phys.\ Rev.\ Fluids} {\bf 11}, 044805.

\bibitem[Kelvin(1887)]{Kelvin1887}
  {\sc Thomson, W. (Lord Kelvin)} 1887
  On ship waves.
  {\it Proc.\ Inst.\ Mech.\ Engrs} {\bf 38}, 409--434.

\bibitem[Voise \& Casas(2010)]{VoiseCasas2010}
  {\sc Voise, J. \& Casas, J.} 2010
  The management of fluid and wave resistances by whirligig beetles.
  {\it J.\ R.\ Soc.\ Interface} {\bf 7}, 343--352.

\end{thebibliography}
\end{document}